\documentclass[aps,prl,twocolumn,showpacs,superscriptaddress,groupedaddress]{revtex4}
\usepackage{graphicx}
\usepackage{dcolumn}
\usepackage{bm}
\usepackage{amssymb}
\usepackage{amsmath}
\usepackage{mathrsfs}
\usepackage{tikz}
\usepackage{tikz-cd}
\usepackage{mathrsfs}
\usepackage{hyperref}

\def\delbar{\bar\partial} 

\newcommand{\Qbar}{\overline{Q}}
\newcommand{\g}{\mathfrak{g}}
\DeclareMathOperator{\Tr}{Tr}

\newcommand{\C}{\mathbb{C}}
\newcommand{\bC}{\mathbb{C}}
\newcommand{\bP}{\mathbb{P}}

\newcommand{\cM}{\mathcal{M}}
\newcommand{\cN}{\mathcal{N}}

\newcommand{\cP}{\mathcal{P}}
\newcommand{\cT}{\mathcal{T}}
\newcommand{\fn}{\mathfrak{n}}
\newcommand{\fh}{\mathfrak{h}}
\newcommand{\bO}{\mathbb{O}}
\newcommand{\cD}{\mathscr{D}}
\newcommand{\hX}{\widehat{X}}

\DeclareMathOperator{\Gr}{Gr}
\DeclareMathOperator{\OG}{OG^{+}}

\begin{document}

\title{Hidden exceptional symmetry in the pure spinor superstring}

\author{R.~Eager}
\affiliation{Kishine Koen, Yokohama, Japan}
\author{G.~Lockhart}
\affiliation{
Institute for Theoretical Physics, University of Amsterdam, Amsterdam, The Netherlands}
\author{E.~Sharpe}
\affiliation{Department of Physics, Virginia Tech, 
Blacksburg, VA, U.S.A.}

\date{\today}

\begin{abstract}
The pure spinor formulation of superstring theory includes an interacting sector of central charge $c_{\lambda}=22$, which can be realized as a curved $\beta\gamma$ system on the cone over the orthogonal Grassmannian $\OG(5,10)$. We find that the spectrum of the $\beta\gamma$ system organizes into representations of the $\mathfrak{g}=\mathfrak{e}_6$ affine algebra at level $-3$, whose  $\mathfrak{so}(10)_{-3}\oplus {\mathfrak u}(1)_{-4}$ subalgebra encodes the rotational and ghost symmetries of the system. As a consequence, the pure spinor partition function decomposes as a sum of affine $\mathfrak{e}_6$ characters. We interpret this as an instance of a more general pattern of enhancements in curved $\beta\gamma$ systems, which also includes the cases $\mathfrak{g}=\mathfrak{so}(8)$ and $\mathfrak{e}_7$, corresponding to target spaces that are cones over the complex Grassmannian $\Gr(2,4)$ and the complex Cayley plane $\mathbb{OP}^2$. We identify these curved $\beta\gamma$ systems with the chiral algebras of certain $2d$ $(0,2)$ CFTs arising from twisted compactification of 4d $\cN=2$ SCFTs on $S^2$.
\end{abstract}

\maketitle

\section{Introduction} 

The pure spinor formalism \cite{Berkovits:2000fe} is a reformulation of superstring theory which has the virtue that it can be quantized while preserving manifest covariance with respect to ten-dimensional super-Poincar\'e symmetry. It therefore provides a powerful approach to quantizing the superstring in curved backgrounds with Ramond--Ramond flux and computing multi-loop scattering amplitudes.
Focusing on the left-movers, the defining feature of this formalism is the presence of a ghost system that is realized in terms of a set of bosonic fields, $\lambda^\alpha,$ transforming in the $\mathbf{16}$ of $\mathfrak{so}(10)$, satisfying the `pure spinor' constraint
\begin{equation}
\lambda^\alpha \gamma^\mu_{\alpha\beta} \lambda^\beta = 0,\qquad \mu=0,\dots,9,
\label{eq:psc}
\end{equation}
and contributing $c_\lambda=22$ to the left central charge. 
In this letter, we will argue that the pure spinor ghost sector possesses a hidden affine $\widehat{\mathfrak{e}}_6$ symmetry at level $-3$, albeit with a choice of stress tensor different from the Sugawara one.  With this choice of stress tensor, only the currents for the $\widehat{\mathfrak{so}}(10)_{-3}\oplus \widehat{\mathfrak{u}}(1)_{-4}$ subalgebra corresponding to rotational and ghost symmetries have conformal dimension $1$. Nevertheless, we find that the ghost sector partition function \cite{Aisaka:2008vw} can be expressed as a linear combination of $(\widehat{\mathfrak{e}}_6)_{-3}$ affine characters:
\begin{equation}
Z = \widehat\chi^{(\widehat{\mathfrak{e}}_6)_{-3}}_{0}-\widehat\chi^{(\widehat{\mathfrak{e}}_6)_{-3}}_{-3\omega_1}.
\end{equation}

To motivate our results, we will start by briefly recalling different known realizations of the ghost system. A convenient realization is as a curved $\beta \gamma$ system on the space $\mathcal{P}$ of 10d pure spinors. A first hint of the enlarged symmetry follows from the work of Levasseur, Smith, and Stafford \cite{MR953165} who found that the space of differential operators on $\mathcal{P}$ enjoys an action of $\mathfrak{e}_6$; see also \cite{brylinski1994minimal} and especially \cite{enright2004resolutions}. In the physics literature, the existence of an $\mathfrak{e}_6$ finite-dimensional Lie algebra action on the zero modes of the pure spinor ghost sector was first observed in \cite{Pioline:2003bk,Pioline:2004xq}.

We will find it enlightening to consider a more general family of $\beta\gamma$ systems whose target spaces, $\widehat{X}_{\mathfrak{g}},$ have enlarged symmetry $\mathfrak{g}= \mathfrak{d}_4 (= \mathfrak{so}(8)), \mathfrak{e}_6$, and $\mathfrak{e}_7$. These varieties can be described as cones over the complex Grassmannian $\Gr(2,4)$, the complex orthogonal Grassmannian $\OG(5,10)$, and the complex Cayley plane $\mathbb{OP}^2$, respectively. An insightful way to realize the target spaces $\widehat{X}_{\mathfrak{g}}$ is as Lagrangian submanifolds of the moduli spaces $\widetilde{\mathcal{M}}_{\mathfrak{g},1}$ of a single centered $\mathfrak{g}$-instanton. These moduli spaces are in turn the Higgs branches of a family of $4d$ $\cN=2$ superconformal field theories (SCFTs) $\mathcal{T}_{\mathfrak{g}}$, whose chiral algebra in the sense of \cite{Beem:2013sza} is the vacuum module of $\widehat{\mathfrak{g}}_{k}$ affine algebra, where $k=-2,-3,-4$ respectively for $\mathfrak{g}= \mathfrak{d}_4, \mathfrak{e}_6$, and $\mathfrak{e}_7$. Applying a topological twist to $\mathbf{\mathcal{T}}_{\mathfrak{g}}$ and reducing on $S^2$ \cite{Cecotti:2015lab}, we will obtain \cite{wip} a set of $2d$ $(0,2)$ CFTs $\mathcal{T}^{(0,2)}_{\mathfrak{g}}$, whose chiral algebras we will identify with the corresponding $\beta\gamma$ system on $\widehat{X}_{\mathfrak{g}}$.
We will present a detailed analysis of these theories in a companion paper \cite{wip}.

The global symmetry of the $\beta\gamma$ systems is a certain maximal subalgebra $\mathfrak{g}_0$ of $\mathfrak{g}$. However, we will see that from the perspective of geometric representation theory, it is natural to expect the entire $\mathfrak{g}$ to act on the states of the theory. This is indeed the case for the theory $\mathbf{\mathcal{T}}^{(0,2)}_{\mathfrak{so}(8)}$, whose chiral algebra was found by Dedushenko and Gukov  \cite{Dedushenko:2017osi} to realize the  $\widehat{\mathfrak{so}}(8)_{-2}$ affine algebra.

We next study how the enlarged symmetry manifests itself in the partition function for $\mathfrak{g}=\mathfrak{so}(8)$ and $\mathfrak{g}=\mathfrak{e}_6$. In both cases, we will find that the partition function can be expressed as a linear combination of two $\widehat{\mathfrak{g}}_{k}$ characters. These results suggest that the chiral algebra of $\mathbf{\mathcal{T}}^{(0,2)}_{\mathfrak{g}}$ receives two types of contributions: one from states arising from the reduction of the $4d$ $\mathcal{N}=2$ chiral algebra, and one capturing contributions from a surface defect of the 4d SCFT
$\cT_{\g}$ that is wrapped along $S^2$.

We will also find an elegant closed form for the partition function of the pure spinor ghost system, written in terms of $\mathfrak{e}_6$ Weyl invariant Jacobi forms, which agrees exactly with the first six energy levels computed in \cite{Aisaka:2008vw}.  An amusing consequence is that the fields, ghosts, anti-fields, and anti-ghosts of ten-dimensional supersymmetric Yang--Mills theory organize into the $\mathbf{27}$ and $\overline{\mathbf{27}}$ of $\mathfrak{e}_6.$\\
\textbf{Note added: } We wish to thank B. Pioline and M. Movshev for bringing references \cite{brylinski1994minimal,Pioline:2003bk,Pioline:2004xq} to our attention, where the existence of an action of the $\mathfrak{e}_6$ finite Lie algebra on the ground states of the pure spinor system was discussed. M. Movshev has informed us of unpublished work \cite{MovshevUnpub} in which the presence of an affine $\mathfrak{e}_6$ symmetry in the pure spinor system has also been studied.

\section{The pure spinor ghost system}

In the pure spinor formalism, the superstring is described by a sigma model consisting of maps $\Sigma \rightarrow \cM$ from the string worldsheet $\Sigma$ into ten-dimensional super-Minkowski space $\cM$ coupled to a ghost system of central charge $c_\lambda=22$. The matter fields on the worldsheet include a set of bosonic fields $x^{\mu}$ in the $\mathbf{10}$ of $\mathfrak{so}(10)$ and, focusing on the left-movers, a set of periodic fermions $ \theta^{\alpha}$ in the $\mathbf{16}$ of $\mathfrak{so}(10)$, along with their conjugate momenta $p_\alpha$, so that the total left-moving central charge $c_L=c_x+c_\theta+c_\lambda=0$.

In the original  \emph{`minimal'} formalism, the ghost sector is captured by a sigma model describing maps $\Sigma \rightarrow \cP$ into the space of ten-dimensional Cartan pure spinors $\cP$, which is parametrized by the bosonic fields $\lambda^\alpha$ satisfying the constraint in equation \eqref{eq:psc}. These fields are accompanied by their conjugate momenta $w_{\alpha}$.
The physical spectrum is given by the cohomology of the nilpotent BRST operator
\begin{equation}
Q = \int \lambda^{\alpha} \bigg(p_{\alpha} + (\gamma^{\mu} \theta_{\alpha}) \partial x^{\mu} - \frac{1}{2} ( \gamma^{\mu} \theta)_{\alpha}(\theta \gamma_{\mu} \partial \theta)\bigg)
\end{equation}
acting on a suitable Hilbert space $\mathcal{H}.$
The Hilbert space can be defined using the curved $\beta\gamma$ system \cite{Nekrasov:2005wg, Aisaka:2008vx,Aisaka:2008vw}. A convenient way to do this is to pass to the {\it non-minimal}  formalism \cite{Berkovits:2005bt}, where physical states are identified with the cohomology of the modified BRST operator
$\Qbar = Q + \delbar,$ where $\delbar$ is a Dolbeault operator acting on $\cP.$  As we will see, the symmetry of the quantum mechanics on $\cP$ is enlarged from $\mathfrak{so}(10)\oplus\mathfrak{u}_1$ to $\mathfrak{e}_6$.  In the next section, we will consider similar spaces
with quantum mechanical symmetry enhancement considered in \cite{MR953165}.
For special target spaces including $\cP$, we will argue that the quantum mechanical enhancement will extend to enhancement 
in the $\beta\gamma$ system.

\section{$\beta\gamma$ systems on complex cones}

Curved $\beta\gamma$ systems  \cite{MR2058353, MR1704283, Witten:2005px, Nekrasov:2005wg}  are two-dimensional sigma models of holomorphic maps $\gamma:\Sigma \rightarrow \hX$ with action
\begin{equation}
S = \frac{1}{2 \pi} \int_\Sigma \beta_i \delbar \gamma^i,
\end{equation}
where, in a given patch, $\gamma_i,\, i=1,\dots,\dim\hX$ serve as local coordinates, $\beta_i$ are $(1,0)$-forms, and 
\begin{equation}
\gamma^i(z)\beta_j(w)\sim\delta^i_j\frac{dw}{z-w}.
\end{equation}
We consider the case where $\hX= \hX_{\mathfrak{g}}$ is one of the varieties constructed by Levasseur, Smith, and Stafford \cite{MR953165}, which is associated to a Lie algebra $\mathfrak{g}$.  Constructing these varieties involves a choice of a minuscule root of $\g$. The minuscule root defines a decomposition of $\g$ of the form
\begin{equation}
\label{eq:grading}
\g = \g_{-1} \oplus \g_0 \oplus \g_1
\end{equation}
where  $\g_0=\mathfrak{u}_1\oplus\mathfrak{s}$, and $\g_1$ is a {\it minuscule} representation $V_{\omega}$ associated to the highest weight $\omega$ of the semi-simple Lie algebra $\mathfrak{s}$ \cite{MR1966752}. Let $G_s$ be the simply connected complex Lie group corresponding to $\mathfrak{s},$ and $P_{\omega}$ be the parabolic subgroup corresponding to $\omega.$ Then, one defines $\hX_{\mathfrak{g}}$ to be the complex cone over the base $\mathbb{P}(\hX_{\g}) = G_s/P_{\omega}.$ The spaces $\hX_{\g}$ have the following homogeneous coordinate ring:
\begin{equation}
\label{eq:hcoord}
\C[\hX_{\g}] \cong \bigoplus_{\ell \ge 0} V_{\ell \omega}.
\end{equation}

We focus on the following cases, where $\mathfrak{g}$ belongs to the Deligne-Cvitanovi\'c exceptional series:
\begin{table}[h!]
\begin{center}
\begin{tabular}{|c|c|c|c|c|c|c|}
\hline
$\g$ & $\omega$ & $\mathfrak{s}$ & $\dim \g_1$ & $\dim_\bC \hX_{\g}$ &  $\mathbb{P}(\hX_{\g})$ &$c_1(\mathbb{P}(\hX_{\g}))$ \\
\hline
$\mathfrak{d}_4$ &  $\omega_4$ & $\mathfrak{a}_{3}$ & 6 & 5 & $\Gr(2,4)$ &4\\
$\mathfrak{e}_6$ & $\omega_1$ & $\mathfrak{d}_5$ & 16 & 11 & $\OG(5,10)$ &8\\
$\mathfrak{e}_7$ & $\omega_7$ & $\mathfrak{e}_6$ & 27 & 17 & $\mathbb{OP}^2$ &12\\
\hline
\end{tabular}
\end{center}
\caption{Relevant varieties.}
\label{tab:minuscule}
\end{table}

\noindent For ${\mathfrak e}_6,$ $\g_0$ is the Lie algebra $\mathfrak{d}_5 \oplus {\mathfrak u}_1$, $V_{\omega}=V_{\omega_4}$ is the spinor representation $\bf{16}$ of $\mathfrak{so}(10),$ and $\hX_\g$ coincides with $\cP$, the space of ten-dimensional pure spinors.

The $\beta\gamma$ system on $\hX_\g$ has central charge $c= 2 \dim_{\C} \hX_\g$ and manifest global symmetry $\g_0=\mathfrak{u}_1\oplus \mathfrak{s}$, where the abelian factor acts by rescaling the cone and $\mathfrak{s}$  acts on the base.  We consider the following partition function:
\begin{equation}
Z_{\g}(t,\mathbf{m}^{\mathfrak{s}},\tau)\!=\!\Tr_{\mathcal{H}} (-1)^F e^{2\pi i \tau H} t^{J-\tfrac{1}{2}a_{\mathfrak{u}_1}}\!\exp(2\pi i \mathbf{m}^{\mathfrak{g}}),
\label{eq:Zpurespinor}
\end{equation}
where $q=e^{2\pi i \tau}$, $t=e^{2\pi i \sigma}$, $F$ is the fermion number, $H$ is the (left) Hamiltonian, $J $ is the $\mathfrak{u}_1$ generator, and $\mathbf{m}^{\mathfrak{s}}\in \mathfrak{h}(\mathfrak{s})_{\mathbb{C}}$ are fugacities for $\mathfrak{s}$. Our definition for the partition function differs from that of \cite{Berkovits:2005hy, Grassi:2005jz, Aisaka:2008vw} by a factor of $t^{-\frac{1}{2}a_{\mathfrak{u}_1}}$, where 
\begin{equation}
a_{\mathfrak{u}_1} = -c_1(\bP(\hX_\g))
\end{equation}
is the $\mathfrak{u}_1$ symmetry anomaly appearing in the operator product expansion (OPE)
\begin{equation}
J(y)T(z) \sim \frac{a_{\mathfrak{u}_1}}{(y-z)^3}+\frac{J(z)}{(y-z)^2}.
\label{eq:anom}
\end{equation}
The $\mathfrak{u}_1$ level, which appears in the OPE
\begin{equation}
J(y)J(z) = \frac{k_{\mathfrak{u}_1}}{(y-z)^2},
\end{equation}
is given in this class of models by
\begin{equation}
k_{\mathfrak{u}_1}=\frac{1}{2}a_{\mathfrak{u}_1}.
\label{eq:anomrel}
\end{equation}
 The $\mathfrak{u}_1$ symmetry anomaly and level can be extracted from the unrefined Hilbert series of $\hX_{g}$ \cite{Berkovits:2005hy}.
 
 The partition function displays the field-antifield symmetry
\begin{equation}
Z_{\g}(t,\mathbf{m}^{\mathfrak{s}},\tau) = (-1)^{\dim_\bC \hX_\g} Z_{\\g}(1/t,-\mathbf{m}^{\mathfrak{s}},\tau)
\end{equation}
and $\star$-conjugation symmetry
\begin{equation}
Z_{\g}(t,\mathbf{m}^{\mathfrak{s}},\tau) = -(q^{\frac{1}{2}}t^{-1})^{\frac{1}{2}c_1(\bP(\hX_\g))} Z_{\g}(q/t,-\mathbf{m}^{\mathfrak{s}},\tau)
\label{eq:star}
\end{equation}
of the $\beta\gamma$ system \cite{Aisaka:2008vx,Aisaka:2008vw}. The ground states contribute
\begin{equation}
q^{-\frac{c}{24}}t^{\frac{1}{2}c_1(\hX_\g)}HS_{\g}
\label{eq:gstates}
\end{equation}
to the partition function, where $HS_{\g} \!\!=\!\! \sum_{\ell=0}^\infty \!t^\ell \chi^{\mathfrak{s}}_{_{V_{\ell \omega}}}\!(\mathbf{m}^{\mathfrak{s}})$ is the refined Hilbert series of $\hX_\g$.

\section{$\beta\gamma$ system from 4d/2d SCFT}

Superconformal field theory (SCFT) provides an additional vantage point from which the curved $\beta\gamma$ systems can be studied. Indeed, the curved $\beta \gamma$ system with target $\hX_\g$ can be identified with the holomorphic twist of a two-dimensional $(0,2)$ sigma model on $\hX_\g$, which also implies equality between the partition function of the former and the elliptic genus of the latter \cite{MR2827826,Costello:2011nq, Gorbounov:2016oia}.

To realize the $(0,2)$ sigma models with the targets $\hX_\g$ listed in table \ref{tab:minuscule}, we begin with a triplet of four-dimensional SCFTs $\cT_{\g}$, where $\g$ denotes the Lie algebra of the flavor symmetry group of $\cT_{\g}$. The theory $\cT_{\mathfrak{d}_4}$ is the $\mathcal{N}=2$ Super-Yang-Mills theory with gauge group $SU(2)$ and four flavors, while $\cT_{\mathfrak{e}_6}$ and $\cT_{\mathfrak{e}_7}$ are the rank-one $\mathfrak{e}_6$ and $\mathfrak{e}_7$ Minahan--Nemeschansky theories \cite{Minahan:1996fg, Minahan:1996cj}. We next perform a partial $N=-1$ topological twist on the $\cN=2$ SCFT along the lines of \cite{Cecotti:2015lab}, and reduce the theory on a two-sphere $S^2$, leading to a two-dimensional theory preserving $(0,2)$ supersymmetry.

Four-dimensional $\cN = 2$ SCFTs have both a Higgs branch and an associated vertex operator algebra (VOA) \cite{Beem:2013sza}. These invariants are intricately related to each other \cite{Beem:2017ooy}. The Higgs branch $Higgs(\cT_{\g})$ is the minimal (non-zero) nilpotent orbit $\bO$ of $\g$, which is also the centered moduli space of one $G$-instanton \cite{MR1072915, MR1649625}, and has complex dimension $2 h^{\vee} - 2.$
Algebraically, the minimal nilpotent orbit is the associated variety of the {\it Joseph ideal} $\mathcal{J}_0$ \cite{MR0404366}.
The spaces $\hX_{\g}$ are Lagrangian submanifolds of $Higgs(\cT_{\g})$. To see this, we first fix a triangular decomposition  $\g = \fn^{-} \oplus \fh \oplus \fn^{+}$  of $\g.$ The irreducible components of the intersection of $\bO$ with $\fn^{+}$ are called {\it minimal orbital varieties}.  They are isotropic subspaces with respect to the Killing form, of dimension $\frac{1}{2} \dim \bO$, hence Lagrangian subvarieties of $\bO$ by theorem 3.3.6 of \cite{MR1433132}, and play an important role in geometric representation theory \cite{MR741942, MR953165, MR939476, MR1604290}.  Smooth orbital varieties of the minimal nilpotent orbit are {\it minuscule varieties}
\cite{fresse:hal-01739780}.  The spaces $\hX_\g$ are minuscule varieties for $G_s.$

\begin{table}[t]
\begin{center}
\begin{tabular}{|c|c|c|c|c|c|c|}
\hline
$\cT_\g$ & $h^{\vee}$& $c_{Sug}$ & $a^{4d}$ & $c^{4d}$ & $h_{min}$ & $c_{eff}$ \\
\hline
$\phantom{\bigg(}\cT_{\mathfrak{d}_4}$& $6$& $-14$  &  $\frac{23}{24}$ & $\frac{7}{6}$ & $-1$ & $10$  \\
$\phantom{x}\cT_{\mathfrak{e}_6}$ & $12$&  $-26$ & $\frac{41}{24}$ & $\frac{13}{6}$& $-2$ & $22$  \\
$\phantom{\bigg(}\cT_{\mathfrak{e}_7}$ & $18$& $-38$& $\frac{59}{24}$ & $\frac{19}{6}$  & $-3$ & $34$ \\
\hline
\end{tabular}
\end{center}
\caption{Properties of the $\cT_\g$ theories.}
\label{tab:4d}
\end{table}

The associated VOA, $V_k(\g),$ is the affine algebra $\widehat{\g}_k$ at level $k = -h^{\vee}/6 - 1,$ where $h^{\vee}$ is the dual Coxeter number of $\g$ \cite{Arakawa:2017fdq}. The holomorphic twist of the $(0,2)$ theory is a chiral theory whose spectrum organizes into representations of the VOA. In particular, from the growth of states, one can argue that the spectrum must include a representation of dimension $h_{min} =\frac{1}{2}(4a^{4d}-5c^{4d})<0$ \cite{Cecotti:2015lab}, where $(a^{4d},c^{4d})$ are the anomaly coefficients of $\cT_{\g}$ listed in Table \ref{tab:4d}. Thus the central charge of the $(0,2)$ theory is shifted from the Sugawara value 
\begin{equation}
c_{Sug} = \frac{\dim \g \, k}{h^\vee+k}
\end{equation}
to the effective central charge
\begin{equation}
c_{eff} = c_{Sug}-24 h_{min} = 2\dim \hX_\g,
\end{equation}
which coincides with the central charge of the $\beta\gamma$ system on $\hX_\g$. The shift in central charge can be traced back to the fact that the stress-energy tensor differs from the Sugawara stress tensor by a correction term \cite{Berkovits:2005hy,Aisaka:2008vx,Dedushenko:2017osi}:
\begin{equation}
T = T_{Sug} + \partial J,
\end{equation}
which gives rise to the $\mathfrak{u}_1$  anomaly of equation \eqref{eq:anom}.
Since the $J(y) T_{Sug}(z)$ OPE has no anomaly, the $\mathfrak{u}_1$ symmetry anomaly and level are proportional, with relation given by equation \eqref{eq:anomrel}.
Similarly, after the modification of the stress-tensor, the currents in $\g_0$ retain conformal dimension 1, while the currents in $\g_1$ and $\g_{-1}$ acquire the new conformal dimensions 2 and 0, respectively.

Altogether, these considerations suggest that the (0,2) theories we constructed flow in the IR to a sigma model on the Lagrangian submanifold $\hX_\g$ of $Higgs(\cT_\g)$, and that their elliptic genus coincides with the partition function of the corresponding $\beta\gamma$ system. Indeed, for $\g=\mathfrak{d}_4,$ the twisted compactification of $\cT_{\mathfrak{d}_4}$ is the Dedushenko-Gukov (0,2) model, which has been argued to flow to a $(0,2)$ sigma model on $\hX_{\mathfrak{d}_4}$ \cite{Dedushenko:2017osi}. We conjecture that an analogous result holds for $\g=\mathfrak{e}_6$ and $\mathfrak{e}_7$ as well.

\section{Symmetry enhancement}

The $\beta \gamma$ system with target $\hX_{\g}$ has a manifest affine $\widehat{\g}_0 \subset \widehat{\g}$ symmetry.
In this section, we argue that in fact the $\beta\gamma$ system enjoys affine $\widehat{\g}$ symmetry.
First, let us review how enhancement to $\g = \g_{-1}\oplus \g_0\oplus \g_1$ occurs in the quantum mechanics on $\hX_{\g}$, a fact which has been studied in \cite{Pioline:2003bk,Pioline:2004xq}. The differential operators, $\cD(\hX_{\g}),$ on $\hX_{\g}$ are equivalent to $U(\g)/\mathcal{J}_0$,
where $U(\g)$ is the universal enveloping algebra of $\g$ and $\mathcal{J}_0$ is the Joseph ideal  \cite{MR953165}. Infinitesimal rotation and dilatation symmetries of $\hX_{\g}$ are generated by differential operators transforming in $\g_0$. Differential operators realizing $\g_{-1}$ and $\g_1$ also have a simple description: those in $\g_{-1}$ correspond to coordinate multiplication, while those in $\g_{1}$ act like generalized special conformal transformations.

Next let us discuss how the identification between $\cD(\hX_\g)$ and $U(\g)/\mathcal{J}_0$ suggests a relationship between the $\beta \gamma$ system on $\hX_{\g}$ and the VOA $V_k(\g).$ On the one hand, the operators realizing the affine $\hat\g_k$ symmetry in the $\beta\gamma$ system, whose explicit construction we defer to future work \cite{wip}, are expected to reduce to differential operators on $\hX_{\g}$ in the limit of quantum mechanics. On the other hand, the {\it chiralization} of the algebra $U(\g)/\mathcal{J}_{\mathcal{W}} \cong \bC \times U(\g)/\mathcal{J}_0$ is the VOA $V_k(\g)$,
where the ideal $\mathcal{J}_{\mathcal{W}}$ of $U(\g)$ is defined in \cite{MR3773273}.
This means that the Zhu algebra of $V_k(\g)$ is $\bC \times U(\g)/\mathcal{J}_0$.
This suggests that we can view the VOA  $V_k(\g)$ as an algebraic analog of the curved $\beta\gamma$ system on $\hX_{\g}.$
The various relations are summarized in the following diagram:
\begin{center}
\begin{tikzcd}
  \beta\gamma \text{ system on } \hX_{\g} \arrow[r, leftrightarrow] \arrow[d, dashed]
    & \text{affine VOA } V_k(\g) \arrow[d, dashed] \\
  \cD(\hX_{\g}) \arrow[r, leftrightarrow, ""]
& U(\g)/\mathcal{J}_0
\end{tikzcd}
\end{center}

The relation between the $\beta\gamma$ system and the twisted $S^2$ compactification of the $\cT_{\g}$ theory provides a further reason to expect the appearance of the affine $\widehat\g_k$ algebra. Indeed, as we have seen in the previous section, the chiral algebra of the resulting $(0,2)$ model provides a representation of the VOA $V_k(\g)$.

In the remainder of this section, we discuss in detail how the symmetry enhancement manifests itself in the partition function in the $\widehat{\mathfrak{so}}(8)_{-2}$ and $(\widehat{\mathfrak e}_6)_{-3}$ cases. Zhu's theorem \cite{MR1317233} relates the classification of simple highest weight $V_k(\g)$-modules to
Joseph's classification \cite{MR1604290} of simple highest weight $U(\g)/\mathcal{J}_0$-modules \cite{MR3773273}.  We use this classification in the following examples.

\subsection{Enhancement to $(\widehat{\mathfrak{d}}_4)_{-2}$ in the $\Gr(2,4)$ cone $\beta\gamma$ system}

We start with the $\beta\gamma$ system on the complex cone over $\Gr(2,4)$. There exist at least two convenient UV descriptions of the corresponding $(0,2)$ sigma model, for which the appearance of an affine $(\widehat{\mathfrak{d}}_4)_{-2}$ algebra was found in \cite{Dedushenko:2017osi}: the first is as a two-dimensional $(0,2)$ $SU(2)$ gauge theory with four fundamental chiral multiplets, which arises directly from the twisted compactification of the 4d $\mathcal{N}=2$ theory $\cT_{\mathfrak{so}(8)}$; the second is as a $(0,2)$ Landau-Ginzburg model consisting of a single Fermi superfield, $\Psi$ and a set of chiral superfields $\Phi$ in the $\wedge^2\mathbf{4}=\mathbf{6}$ representation of $SU(4)$, coupled via a $J$-type superpotential interaction term $J=\Psi \text{Pf}(\Phi)$. 

Classically, the vacuum moduli space of this model is
a quadric in $\C^6$, specifically the affine cone over $G(2,4)$, which is
the closure of $\hX_{\mathfrak{so}(8)}$.
Quantum corrections will modify this picture in the interior. As we will discuss
in \cite{wip}, in analogy with the pure spinor case, we conjecture \footnote{ 
One trivial consistency check is that the Dedushenko-Gukov model behaves like a GLSM
for a smooth target, in that it does not exhibit pathologies; removing the singular
vertex of the affine cone 
is one way to ensure that the corresponding sigma model has a smooth target.
As another, less trivial, consistency check, there is a short argument due to Ron Donagi that
ch$_2(T \hX_{\mathfrak{so}(8)})$ vanishes, much as ch$_2$ vanishes for the cone over $\OG(5,10)$ in
the pure spinor case \cite{Nekrasov:2005wg}. Briefly, standard exact sequences that express
the K theory class of $T \hX_{\mathfrak{so}(8)}$ can be expressed
in terms of ${\cal O}$, ${\cal O}(1)$, but the latter is trivial since the vertex has  
been removed, suggesting that all Chern classes of $T \hX_{\mathfrak{so}(8)}$ vanish.
}
that quantum
corrections move the singular vertex of the affine cone infinitely far away,
realizing a two-dimensional $(0,2)$ sigma model on $\hX_{\mathfrak{so}(8)}$.

For this theory, $\g_0=\mathfrak{u}_1\oplus \mathfrak{a}_3$ is the manifest global symmetry, while $\mathfrak{g}_{-1}$ = $\mathfrak{g}_1$ is the $V_{\omega_2} = \mathbf{6}$ representation of $\mathfrak{a}_3$. The partition function can be computed straightforwardly, either 
as the elliptic genus of the $(0,2)$ SQCD theory \cite{Putrov:2015jpa} following \cite{Gadde:2013ftv, Benini:2013nda, Benini:2013xpa}, or directly as the partition function of the curved $\beta\gamma$ system on $\Gr(2,4)$ \cite{Aisaka:2008vx}. It is given by:
\begin{equation}
Z_{\mathfrak{d}_4}(t,\mathbf{m}^{\mathfrak{a}_3},\tau) = \frac{i \,\eta(\tau)^5\theta_1(2\sigma,\tau)}{\prod_{w\in \mathbf 6}\theta_1(\sigma+(\mathbf{m}^{\mathfrak{a}_3},w),\tau)}.
\end{equation}
The product in the denominator is over the weights in the $\mathbf{6}$ of $\mathfrak{a}_3$. We now proceed to express the partition function in terms of $(\widehat{\mathfrak{d}}_4)_{-2}$ characters. The embedding of $\mathfrak{u}_1\oplus\mathfrak{a}_3$ into $\mathfrak{d}_4$ implies the following mapping of parameters:
\begin{align*}
\mathbf{m}^{\mathfrak{d}_4}_i&=\mathbf{m}^{\mathfrak{a}_3}_i \qquad\text{for } i=1,2,3;\\
\mathbf{m}^{\mathfrak{d}_4}_4&=\sigma-\frac{\mathbf{m}^{\mathfrak{a}_3}_1}{2}-\mathbf{m}^{\mathfrak{a}_3}_2-\frac{\mathbf{m}^{\mathfrak{a}_3}_3}{2}.
\end{align*}
The algebra $(\widehat{\mathfrak{d}}_4)_{-2}$ is known to possess four irreducible highest weight representations, corresponding to the following choices of highest weight \cite{perse2013note, MR3773273}:
\begin{equation}
0,\,-2\omega_1,\,-2\omega_3,\,-2\omega_4.
\end{equation}
The three non-vacuum representations are related by triality. While each of these highest weights is not dominant, it is still the case that there exists a unique dominant weight $\Lambda $ in the shifted Weyl orbit of the highest weight. As a consequence \cite{MR560412,MR573434}, the corresponding affine characters are determined in terms of Kazhdan-Lusztig polynomials \cite{MR560412,MR573434}. By an explicit computation, we find that the partition function is given by a sum of two characters:
\begin{equation}
Z_{\mathfrak{d}_4}(t,\mathbf{m}^{\mathfrak{a}_3},\tau) = \widehat{\chi}^{(\widehat{\mathfrak{d}}_4)_{-2}}_{0}(\mathbf{m}^{\mathfrak{d}_4},\tau)-\widehat{\chi}^{(\widehat{\mathfrak{d}}_4)_{-2}}_{-2\omega_4}(\mathbf{m}^{\mathfrak{d}_4},\tau).
\end{equation}
The vacuum character has the following $q$-expansion:
\begin{equation}
\widehat{\chi}^{(\widehat{\mathfrak{d}}_4)_{-2}}_{0} = q^{\frac{14}{24}}(1+\chi^{\mathfrak{d}_4}_\mathbf{28}\,q+(\chi^{\mathfrak{d}_4}_\mathbf{300}+\chi^{\mathfrak{d}_4}_\mathbf{28}+1)\,q^2+\dots).
\end{equation}
The non-vacuum character has conformal dimension $h=-1$, consistent with table \ref{tab:4d}. As noted in \cite{Beem:2017ooy}, it has the property that
an infinite number of states appear at each energy level. In particular, its lowest energy level component, expressed in $\mathfrak{u}_1\oplus\mathfrak{a}_3$ fugacities, has the following series expansion:
\begin{equation}
-q^{\frac{14}{24}-1}t^2\sum_{\ell=0}^\infty \!t^\ell \chi^{\mathfrak{a}_3}_{_{V_{\ell \omega_2}}}\!(\mathbf{m}^{\mathfrak{a}_3}),
\end{equation}
which encodes the ground states of the partition function of the $\beta\gamma$ system, equation \eqref{eq:gstates}. At the next energy level, one finds the following contributions:
\begin{equation}
q^{\frac{14}{24}}\bigg(2-t^2(\chi^{\mathfrak{a}_3}_\mathbf{15}+1)-t^2(\chi^{\mathfrak{a}_3}_\mathbf{15}+1)-t^3(\chi^{\mathfrak{a}_3}_\mathbf{64}+2\chi^{\mathfrak{a}_3}_\mathbf{6})+\dots\bigg).
\end{equation}
Interestingly, it appears natural to define the following new combination of characters:
\begin{equation}
\widehat{\xi}^{(\widehat{\mathfrak{d}}_4)_{-2}}_{-2\omega_4}(\mathbf{m}^{\mathfrak{d}_4},\tau) = -\widehat{\chi}^{(\widehat{\mathfrak{d}}_4)_{-2}}_{-2\omega_4}(\mathbf{m}^{\mathfrak{d}_4},\tau)+2\widehat{\chi}^{(\widehat{\mathfrak{d}}_4)_{-2}}_{0}(\mathbf{m}^{\mathfrak{d}_4},\tau),
\end{equation}
in terms of which
\begin{equation}
Z_{\mathfrak{d}_4}(t,\mathbf{m}^{\mathfrak{a}_3},\tau) = -\widehat{\chi}^{(\widehat{\mathfrak{d}}_4)_{-2}}_{0}(\mathbf{m}^{\mathfrak{d}_4},\tau)+\widehat{\xi}^{(\widehat{\mathfrak{d}}_4)_{-2}}_{-2\omega_4}(\mathbf{m}^{\mathfrak{d}_4},\tau).
\end{equation}
The $(\widehat{\mathfrak{d}}_4)_{-2}$ characters, once expressed in terms of $\mathfrak{u}_1\oplus\mathfrak{a}_3$ fugacities, satisfy the following simple relation:
\begin{equation}
\frac{t^2}{q}\widehat{\chi}^{(\widehat{\mathfrak{d}}_4)_{-2}}_{0}(\tau\!-\!\sigma,\mathbf{m}^{\mathfrak{a}_3},\tau)\!=\!\widehat{\xi}^{(\widehat{\mathfrak{d}}_4)_{-2}}_{0}(\sigma,\mathbf{m}^{\mathfrak{a}_3},\tau),
\end{equation}
which guarantees that the partition function obeys the $\star$-conjugation symmetry of equation \eqref{eq:star}.

\subsection{Enhancement  to $(\widehat{\mathfrak{e}}_6)_{-3}$ in the pure spinor $\beta\gamma$ system}

We now turn to the pure spinor $\beta\gamma$ system and discuss how the affine $(\widehat{\mathfrak{e}}_6)_{-3}$ symmetry manifests itself at the level of the partition function.  The partition function has been computed up to energy level five by fixed point techniques in \cite{Aisaka:2008vx}, and an all-order expression in the $\mathbf{m}^{\mathfrak{d}_5}\to 0$ limit was found in \cite{Movshev:2016xze} using local algebra \cite{Movshev:2015vmb}. In what follows, we will be able to give a very simple closed form for the partition function for arbitrary values of $\mathbf{m}^{\mathfrak{e}_6}$ fugacities.

For this theory, $\mathfrak{g}_0$ is the $\mathfrak{u}_1\oplus\mathfrak{d}_5$ ghost and rotational symmetry of the pure spinor ghost system. The components $\mathfrak{g}_{-1}$ and $\mathfrak{g}_{1}$ correspond to the $\mathbf{16}$ and $\mathbf{\overline{16}}$ representations of $\mathfrak{d}_5,$ respectively. We begin by discussing the $\mathfrak{e}_6\to \mathfrak{u}_1\oplus\mathfrak{d}_5$ branching rules. The realization of the space $\mathcal{P}$ of pure spinors as the variety $\widehat{X}_{\mathfrak{e}_6}$  implies the following mapping of parameters between ${\mathfrak e}_6$ and ${\mathfrak u}_1\oplus \mathfrak{d}_5$:
\begin{equation*}
\begin{pmatrix}
m^{{\mathfrak e}_6}_1\\
m^{{\mathfrak e}_6}_2\\
m^{{\mathfrak e}_6}_3\\
m^{{\mathfrak e}_6}_4\\
m^{{\mathfrak e}_6}_5\\
m^{{\mathfrak e}_6}_6
\end{pmatrix}
=
\begin{pmatrix}
-3&-\frac{1}{2}&-1&-\frac{3}{2}&-\frac{3}{4}&-\frac{5}{4}\\
0&0&0&0&1&0\\
0&0&0&0&0&1\\
0&0&0&1&0&0\\
0&0&1&0&0&0\\
0&1&0&0&0&0
\end{pmatrix}
\cdot
\begin{pmatrix}
m^{{\mathfrak u}_1}\\
m^{\mathfrak{d}_5}_1\\
m^{\mathfrak{d}_5}_2\\
m^{\mathfrak{d}_5}_3\\
m^{\mathfrak{d}_5}_4\\
m^{\mathfrak{d}_5}_5
\end{pmatrix}
\end{equation*}
(see appendix A for our conventions). At the level of representations, the $\mathbf{27}$ and adjoint of ${\mathfrak e}_6$ decompose as follows, where the subscript denotes $\mathfrak{u}_{1}$ charge:
\begin{eqnarray*}
\mathbf{27}&\to&\mathbf{1}_{-4}+\mathbf{16}_{-1}+\mathbf{10}_2,\\
\mathbf{78} &\to& \mathbf{\overline{16}}_{-3}+\mathbf{1}_0+\mathbf{45}_0+\mathbf{16}_{3}.
\end{eqnarray*}
Furthermore, to match the pure spinor formalism conventions, in equation~(\ref{eq:Zpurespinor}) we must set 
\begin{equation}
t=e^{2\pi i\cdot(-3\, m^{{\mathfrak u}_1})}.
\end{equation}

The algebra $(\widehat{\mathfrak{e}}_6)_{-3}$ possesses a finite number of irreducible modules \cite{MR3773273} corresponding to the highest weights
\begin{equation}
0,\,-3\omega_1,\,-3\omega_6,\,\omega_1-2\omega_3,\,\omega_6-2\omega_5,\,-2\omega_2,\, -\omega_4.
\end{equation}
The non-vacuum representations have conformal dimension  $-2$ which equals the value of $h_{min}$ given in table \ref{tab:4d}. We find that the pure spinor partition function is given by the following combination of $(\widehat{\mathfrak{e}}_6)_{-3}$ characters:
\begin{equation}
Z_{\mathfrak{e}_6}(t,\mathbf{m}^{\mathfrak{a}_3},\tau) = \widehat{\chi}^{(\widehat{\mathfrak{e}}_6)_{-3}}_{0}(\mathbf{m}^{\mathfrak{e}_6},\tau)-\widehat{\chi}^{(\widehat{\mathfrak{e}}_6)_{-3}}_{-3\omega_1}(\mathbf{m}^{\mathfrak{e}_6},\tau).
\end{equation}
The lowest energy component of the non-vacuum term is the Hilbert series of the \emph{Wallach representation} of the $\mathfrak{e}_6$ finite-dimensional Lie algebra \cite{enright2004resolutions} corresponding to highest weight $-3\omega_1$, up to an overall factor of $-q^{\frac{26}{24}-2}t^4$. Expressed in terms of $\mathfrak{u}_1\oplus \mathfrak{d}_5$ fugacities, it is given by
\begin{equation}
-q^{-\frac{22}{24}}t^4\sum_{\ell=0}^\infty \!t^\ell \chi^{\mathfrak{d}_5}_{_{V_{\ell \omega_5}}}\!(\mathbf{m}^{\mathfrak{d}_5}),
\label{eq:e6gs}
\end{equation}
in agreement with the Hilbert series of the cone over the orthogonal Grassmannian $\OG(5,10)$, which is the space of pure spinors in ten dimensions.

Again, it appears natural to define the following combination of characters:
\begin{equation}
\widehat{\xi}^{(\widehat{\mathfrak{e}}_6)_{-3}}_{-3\omega_1}(\mathbf{m}^{\mathfrak{e}_6},\tau) = -\widehat{\chi}^{(\widehat{\mathfrak{e}}_6)_{-3}}_{-3\omega_1}(\mathbf{m}^{\mathfrak{e}_6},\tau)+2\widehat{\chi}^{(\widehat{\mathfrak{e}}_6)_{-3}}_{0}(\mathbf{m}^{\mathfrak{e}_6},\tau),
\end{equation}
in terms of which
\begin{equation}
Z_{\mathfrak{e}_6}(t,\mathbf{m}^{\mathfrak{d}_5},\tau) = -\widehat{\chi}^{(\widehat{\mathfrak{e}}_6)_{-3}}_{0}(\mathbf{m}^{\mathfrak{e}_6},\tau)+\widehat{\xi}^{(\widehat{\mathfrak{e}}_6)_{-3}}_{-3\omega_1}(\mathbf{m}^{\mathfrak{e}_6},\tau).
\label{eq:e6char}
\end{equation}
The $(\widehat{\mathfrak{e}}_6)_{-3}$ characters, expressed in terms of $\mathfrak{u}_1\oplus\mathfrak{d}_5$ fugacities, satisfy the following relation:
\begin{equation}
\frac{t^4}{q^2}\widehat{\chi}^{(\widehat{\mathfrak{e}}_6)_{-3}}_{0}(\tau\!-\!\sigma,\mathbf{m}^{\mathfrak{d}_5},\tau)\!=\!\widehat{\xi}^{\,(\widehat{\mathfrak{e}}_6)_{-3}}_{0}(\sigma,\mathbf{m}^{\mathfrak{d}_5},\tau),
\end{equation}
which guarantees that the partition function obeys the $\star$-conjugation symmetry described by equation \eqref{eq:star}. The significance of writing the partition function as in equation \eqref{eq:e6char} is that the $\widehat{\xi}^{(\widehat{\mathfrak{e}}_6)_{-3}}_{-3\omega_1}$ character captures the contribution of the globally defined operators, which are identified with the zeroth cohomology $H^0(\bar\partial)$, while $\widehat{\chi}^{(\widehat{\mathfrak{e}}_6)_{-3}}_{0}$ captures the contribution from the `missing states' in the Hilbert space of the pure spinor system, which are built out of the $b$-ghost and correspond to $H^3(\bar\partial)$.

We also find that the pure spinor partition function, written in terms of $\mathfrak{e}_6$ fugacities, can be written in the following very simple closed form:
\begin{equation}
Z_{\mathfrak{e}_6}(\mathbf{m}^{{\mathfrak e}_6},\tau) = \frac{(2i)^{-1}\alpha_{-5,1}(\mathbf{\widetilde{m}^{{\mathfrak e}_6}},\tau)}{\eta(\tau)^{22}\prod_{j=1}^{16}\varphi_{-1,1/2}((\mathbf{m}^{{\mathfrak e}_6},\alpha^\vee_{\mathbf{16},j}),\tau)}.
\label{eq:pse6}
\end{equation}
The product in the denominator is over the subset $\alpha^\vee_{\mathbf{16}}$ of coroots of $\mathfrak{e}_6$ that belong to $\mathfrak{g}_{-1}=\mathbf{16}$ under the grading in equation \eqref{eq:grading}. On the other hand, the numerator is given in terms of the Weyl[$\mathfrak{e}_6$]-invariant Jacobi form $\alpha_{-5,1}^{{\mathfrak e}_6}(\mathbf{m}^{{\mathfrak e}_6},\tau)$ (see appendix B), with the following subtlety: the argument $\mathbf{m}^{{\mathfrak e}_6}$ is replaced by the \emph{shifted} $\mathfrak{e}_6$ fugacities $\widetilde{\mathbf{m}}=\sum_i\widetilde{m}^{{\mathfrak e}_6}_i \omega_i$, where $\widetilde{m}^{{\mathfrak e}_6}_i=m_i$ for $i=2,\dots,6$, but
\begin{equation}
\widetilde{m}^{\mathfrak{e}_6}_1 = -3m_1-3m_2-5m_3-6m_4-4m_5-2m_6.
\label{eq:mshift}
\end{equation}
Under ${\mathfrak e}_6\to \mathfrak{d}_5\oplus {\mathfrak u}_1,$ this shift corresponds to setting $m^{{\mathfrak u}_1}\to -3 m^{{\mathfrak u}_1}$, while keeping $\mathbf{m}^{\mathfrak{d}_5}$ invariant.

Using the modular transformation properties of the numerator (and taking into account the shift \eqref{eq:mshift}), one finds that under $\tau\to -1/\tau$ $Z_{\mathfrak{e}_6}$ transforms as follows:
\begin{equation*}
Z_{\mathfrak{e}_6}\left(\frac{\mathbf{m}^{\mathfrak{e}_6}}{\tau},-\frac{1}{\tau}\right) = \exp\left(-3\frac{\pi\, i}{\tau}(\mathbf{m}^{\mathfrak{e}_6},\mathbf{m}^{\mathfrak{e}_6})\right)Z_{\mathfrak{e}_6}\!\left(\mathbf{m}^{\mathfrak{e}_6},\tau\right),
\end{equation*}
where the phase factor is consistent with the occurrence of the $(\widehat{\mathfrak{e}}_6)_{-3}$ algebra \cite{DelZotto:2018tcj}.

It is now straightforward to express the partition function \eqref{eq:pse6} in terms of the pure spinor fugacities $t $ and $\mathbf{m}^{\mathfrak{d}_5}$ via equations \eqref{eq:thetae616}; after doing so, we find an exact match with \cite{Aisaka:2008vw}, 
where $Z_{\mathfrak{e}_6}$ was computed up to the fifth energy level by fixed point techniques.

Rewriting the partition function as
\begin{equation}
Z_{\mathfrak{e}_6}(\mathbf{m}^{\mathfrak{e}_6},\tau) =\frac{\widehat{\chi}^{(\widehat{\mathfrak{e}}_6)_1}_{\omega_1}(\mathbf{m}^{\mathfrak{e}_6},\tau)-\widehat{\chi}^{(\widehat{\mathfrak{e}}_6)_1}_{\omega_6}(\mathbf{m}^{\mathfrak{e}_6},\tau)}{\prod_{i=1}^{16}\eta(\tau)^{-1}\theta_1((\mathbf{m},\alpha^\vee_{\mathbf{16},i}),\tau)},
\end{equation}
where $\widehat\chi^{(\widehat{\mathfrak{e}}_6)_{1}}_{\omega_{1,6}} = \Theta_{\omega_{1,6}}^{\mathfrak{e}_6}/\eta^6$ are level 1 $\mathfrak{e}_6$ affine characters, suggests a possible alternative interpretation as a level 1 $\mathfrak{e}_6$ sector coupled to $16$ complex bosons; we do not pursue this direction further in this letter.

\section{Conclusions}

We have found that the states in the $\beta\gamma$ system with target $\hX_{\g}$ organize into a direct sum of irreducible modules of an affine $\widehat{\g}$ symmetry algebra.
When the target is the space of ten-dimensional pure spinors, $\cP,$ the symmetry algebra is $(\widehat{\mathfrak{e}}_6)_{-3}$.
This knowledge led us to find a compact closed form expression for the full partition function of the ghost system
of the pure-spinor superstring in equation \eqref{eq:pse6}.
We leave it to future work to study possible implications for the computation of operator product expansions and scattering amplitudes in superstring theory.

While we have given several arguments for the appearance of the $\widehat{\g}_k$ symmetry,
it should be possible to explicitly realize its generators in the curved $\beta\gamma$ system.
We have focused on three different smooth targets for the $\beta\gamma$ system.
It would also be natural to extend our analysis to other targets, possibly with singularities.

We have seen that the appearance of $\widehat{\g}_k$ symmetry
has a natural explanation from the perspective of the four-dimensional SCFT, $\cT_{\g},$ dimensionally reduced on $S^2.$
This also explains the appearance of the vacuum module of $\widehat{\g}$ in the elliptic genus. It remains however to explain the occurrence of a second module. A possible hint is that the unflavored limit of the vacuum character of the VOA $V(\g_k)$, for $\g$ belonging to the Deligne-Cvitanovi\'c exceptional series, satisfies a second order {\it linear modular differential equation} \cite{Arakawa:2016hkg, Beem:2017ooy}; the other solution has been conjectured by Beem and Rastelli to arise from a surface operator in the $\mathcal{T}_{\g}$ theory. This suggests an interpretation of the second module in the $(0,2)$ theory as originating from a surface defect wrapping the $S^2.$ We plan to return to this issue in a separate work \cite{wip}.

We would like to thank C.~Beem, T.~Creutzig, J.~Distler, R.~Donagi, I.~Melnikov, M.~Movshev, B.~Pioline, I.~Saberi, and J.~Song for valuable discussions and correspondence.  R.E. thanks the Korean Institute for Advanced Study for hospitality. The work of G.L. is supported by ERC starting grant H2020 ERC StG \#640159. 
The work of E.S. is partially supported by NSF grant PHY-1720321.

\appendix

\section{Appendix A: Lie algebras.}

%
\begin{figure}
\begin{picture}(120,60)
\put(20,40){$\mathfrak{a}_3$:}
\put(50,55){\circle{6}}
\put(56,58){${\!\!\!\!\!\!}^{\mathbf{4}}$}
\put(50, 45){${\!\!\!\!\>}_{\omega_1}$}
\put(53,55){\line(1,0){24}}
\put(80,55){\circle{6}}
\put(77,58){${\hspace{.01in}}^{\mathbf{6}}$}
\put(80, 45){${\!\!\!\!\>}_{\omega_2}$}
\put(83,55){\line(1,0){24}}
\put(110,55){\circle{6}}
\put(115,58){${\!\!\!\!}^{\overline{\mathbf{4}}}$}
\put(110, 45){${\!\!\!\!\>}_{\omega_3}$}
\put(80,26){\circle{6}}
\put(70,21){${\hspace{.01in}}^{\mathbf{1}}$}
\put(90,26){${\!\!\!\!\>}_{\omega_0}$}
\put(82.5,28){\line(1,1){25}}
\put(77.5,28){\line(-1,1){25}}
\end{picture}
\begin{picture}(120,60)
\put(20,40){$\mathfrak{d}_4$:}
\put(50,30){\circle{6}}
\put(50,33){${\!\!\!\!\!\!}^{\mathbf{1}}$}
\put(50, 20){${\!\!\!\!\>}_{\omega_0}$}
\put(53,30){\line(1,0){24}}
\put(80,30){\circle{6}}
\put(80,33){${\hspace{.01in}}^{\mathbf{28}}$}
\put(86, 20){${\!\!\!\!\>}_{\omega_2}$}
\put(83,30){\line(1,0){24}}
\put(110,30){\circle{6}}
\put(110,33){${\!\!\!\!}^{\mathbf{8_s}}$}
\put(110, 20){${\!\!\!\!\>}_{\omega_4}$}
\put(80,33){\line(0,1){24}}
\put(80,60){\circle{6}}
\put(82,66){${\hspace{-1.0em}}_{\mathbf{8_c}}$}
\put(80,64){${\hspace{0.4em}}_{\omega_3}$}
\put(80,3){\line(0,1){24}}
\put(80,00){\circle{6}}
\put(77,0){${\hspace{-1.0em}}_{\mathbf{8_v}}$}
\put(80,0){${\hspace{0.4em}}_{\omega_1}$}
\end{picture}
\begin{picture}(120,60)
\put(-40,40){$\mathfrak{d}_5$:}
\put(20,30){\circle{6}}
\put(20,33){${\!\!\!}^\mathbf{1}$}
\put(20, 20){${\!\!\!\!\>}_{\omega_0}$}
\put(23,30){\line(1,0){24}}
\put(50,33){\line(0,1){24}}
\put(50,60){\circle{6}}
\put(50,66){${\hspace{-1.0em}}_{\mathbf{10}}$}
\put(50,64){${\hspace{0.4em}}_{\omega_1}$}
\put(50,30){\circle{6}}
\put(50,33){${\!\!\!\!\!\!}^{\mathbf{45}}$}
\put(50, 20){${\!\!\!\!\>}_{\omega_2}$}
\put(53,30){\line(1,0){24}}
\put(80,30){\circle{6}}
\put(80,33){${\hspace{.01in}}^{\mathbf{120}}$}
\put(80, 20){${\!\!\!\!\>}_{\omega_3}$}
\put(83,30){\line(1,0){24}}
\put(110,30){\circle{6}}
\put(110,33){${\!\!\!\!}^{\mathbf{16}}$}
\put(110, 20){${\!\!\!\!\>}_{\omega_5}$}
\put(80,33){\line(0,1){24}}
\put(80,60){\circle{6}}
\put(80,66){${\hspace{-1.0em}}_{\mathbf{\overline{16}}}$}
\put(80,64){${\hspace{0.4em}}_{\omega_4}$}
\end{picture}
\begin{picture}(150,90)
\put(-25,60){$\mathfrak{e}_6$:}
\put(20,30){\circle{6}}
\put(20,33){${\!\!\!}^\mathbf{27}$}
\put(20, 20){${\!\!\!\!\>}_{\omega_1}$}
\put(23,30){\line(1,0){24}}
\put(50,30){\circle{6}}
\put(50,33){${\!\!\!\!}^{\mathbf{\overline{351\!\!}}}$}
\put(50, 20){${\!\!\!\!\>}_{\omega_3}$}
\put(53,30){\line(1,0){24}}
\put(80,30){\circle{6}}
\put(80,33){${\hspace{.01in}}^{\mathbf{2925}}$}
\put(80, 20){${\!\!\!\!\>}_{\omega_4}$}
\put(83,30){\line(1,0){24}}
\put(110,30){\circle{6}}
\put(110,33){${\!\!\!\!}^{\mathbf{351}}$}
\put(110, 20){${\!\!\!\!\>}_{\omega_5}$}
\put(113,30){\line(1,0){24}}
\put(140,30){\circle{6}}
\put(140,33){${\!\!}^{\mathbf{\overline{27\!}}}$}
\put(140, 20){${\!\!\!\!\>}_{\omega_6}$}
\put(80,33){\line(0,1){24}}
\put(80,60){\circle{6}}
\put(80,66){${\hspace{-1.0em}}_{\mathbf{78}}$}
\put(80,64){${\hspace{0.4em}}_{\omega_2}$}
\put(80,63){\line(0,1){24}}
\put(80,90){\circle{6}}
\put(80,96){${\hspace{-.7em}}_{\mathbf{1}}$}
\put(80,94){${\hspace{0.4em}}_{\omega_0}$}
\end{picture}
\caption{Dynkin diagrams showing our labeling conventions.}
\label{fig:e6}
\end{figure}
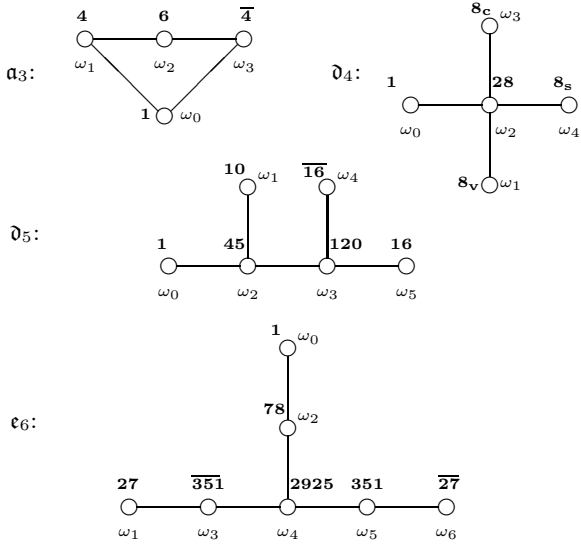
Given a Lie algebra $\mathfrak{g}$, let $\mathfrak{h}(\mathfrak{g})$ be its Cartan subalgebra, $\Delta^{\mathfrak{g}}$ its root lattice, $\alpha_i$, $i=1,\dots,\text{rank}(\mathfrak{g})$ a choice of simple roots, and $\alpha^\vee_i$ the corresponding coroots. The fundamental weights $\omega_i$ are defined by
\begin{equation}
(\omega_i,\alpha^\vee_j)= \delta_{ij},
\end{equation}
where $(\cdot,\cdot)$ is the invariant bilinear form on $\mathfrak{h}(\mathfrak{g})_{\mathbb{C}}$, normalized such that $(\alpha,\alpha)=2$ for the long roots. We adopt Bourbaki's numbering convention for the $\omega_i$. We denote the irreducible representations associated to the highest weight $\omega$ either by $V_\omega$ or by its dimension, following the conventions of \cite{Feger15} (see e.g. figure \ref{fig:e6}).
The character of a highest weight representation $\mathbf{R}$ of $\mathfrak{g}$ is given by
\begin{equation}
\chi_{\mathbf{R}}(\mathbf{m}^{\mathfrak{g}}) = \sum_{w\in \mathbf{R}} \exp\big(2\pi i (\mathbf{m}^{\mathfrak{g}},w)\big),
\end{equation}
where  $\mathbf{m}^{\mathfrak{g}} = \sum_i m^{\mathfrak{g}}_i \omega_i \in\mathfrak{h}^*_{\mathbb{C}}$. The character of a $\mathfrak{g}=\mathfrak{u}_1$ representation of charge $k$ is just $e^{2\pi i k m^{\mathfrak{u}_1}}$.
For an affine Lie algebra, we denote a highest weight representation simply by the finite part $\omega =\sum_i n_i\omega_i$ of its highest weight $(\omega,k,n)$. We denote by $\widehat\chi^{\widehat{\mathfrak{g}}_k}_\omega(\mathbf{m}^{\mathfrak{g}},\tau)$ the corresponding affine character.

\section{Appendix B: Modular and Jacobi forms}

The Dedekind $\eta$ function is defined as follows:
\begin{equation}
\eta(\tau) = q^{1/24}(q;q)_\infty = q^{1/24}\prod_{k=1}^\infty(1-q^k).
\end{equation}
The Jacobi theta functions are given by:
\begin{eqnarray*}
\theta_1(\zeta,\tau)\!\!&=&\! i\!\!\sum_{n\in\mathbb{Z}+\frac{1}{2}}\!(-1)^n z^{n}q^{\frac{n^2}{2}}\!,\,\,\theta_2(\zeta,\tau)\!=\!\!\!\sum_{n\in\mathbb{Z}+\frac{1}{2}}\!z^{n}q^{\frac{n^2}{2}},\\
\theta_3(\zeta,\tau)\!\!&=&\!\! \sum_{n\in\mathbb{Z}}z^{n}q^{\frac{1}{2}n^2},\quad\theta_4(\zeta,\tau) = \sum_{n\in\mathbb{Z}}(-1)^nz^{n}q^{\frac{1}{2}n^2},
\end{eqnarray*}
where $z=e^{2\pi i \zeta}$. Closely related is the weak Jacobi form of weight $-1$ and index $\frac{1}{2}$
\begin{equation}
\varphi_{-1,1/2}(\zeta,\tau) = \frac{\theta_1(\zeta,\tau)}{\eta(\tau)^3}.
\end{equation}
We also make use of Weyl-invariant weak Jacobi forms \cite{MR1163219, MR1249221, Sakai:2017ihc, DelZotto:2017mee}. Under a modular transformation, a Weyl[$\mathfrak{g}$]-invariant weak Jacobi form $\varphi_{w,n}: \mathfrak{h}(\mathfrak{g})\times \mathbb{H}\to\mathbb{C}$ of weight $w$ and index $n$ transforms as:
\begin{equation}
\varphi_{w,n}\!\!\left(\!\frac{\mathbf{z}}{c\tau\!+\!d},\!\frac{a\tau\!+\!b}{c\tau\!+\!d}\!\right)\!\! = \!(c\tau\!+\!d)^{\!w}\! \exp\!\!\left(\!\frac{\pi\, i\, n\,c}{c\tau\!+\!d}(\mathbf{z},\mathbf{z})\!\!\right)\!\!\varphi_{w,n}(\mathbf{z},\tau),
\end{equation}
while for $s\in\text{Weyl}[\mathfrak{g}]$
\begin{equation}
 \varphi_{w,n}(s\mathbf{z},\tau)=\varphi_{w,n}(\mathbf{z},\tau).
\end{equation}

Denote by  $J_{w,n}^{\mathfrak{g}}$ the vector space of Weyl[$\mathfrak{g}$]-invariant weak Jacobi forms of weight $w$ and index $n$. For $\mathfrak{g}\neq {\mathfrak e}_8$, the bigraded ring $J_{*,*}^{\mathfrak{g}} = \bigoplus_{w,n}J_{w,n}^{\mathfrak{g}}$ is a polynomial ring over the ring of $SL(2,\mathbb{Z})$ modular forms, which is known to be generated by $\text{rk}(\mathfrak{g})+1$ independent forms $\alpha^{\mathfrak{g}}_{w,n}$ of specified weight and index. For $\mathfrak{g}={\mathfrak e}_6$, the seven generators
\begin{equation*}
\alpha^{\mathfrak{e}_6}_{0,1},\quad\!\alpha^{\mathfrak{e}_6}_{-2,1},\quad\!\alpha^{\mathfrak{e}_6}_{-5,1},\quad\!\alpha^{\mathfrak{e}_6}_{-6,2},\quad\!\alpha^{\mathfrak{e}_6}_{-8,2},\quad\!\alpha^{\mathfrak{e}_6}_{-9,2},\quad\!\alpha^{\mathfrak{e}_6}_{-12,3}.
\end{equation*}
have been constructed in \cite{MR1249221, Sakai:2017ihc}. In this letter, we make use of the unique Weyl[$\mathfrak{e}_6$]-invariant weak Jacobi form of weight $-5$ and index $1$,
\begin{equation}
\label{eq:alpha}
\alpha^{\mathfrak{e}_6}_{-5,1}(\mathbf{m}^{\mathfrak{e}_6},\tau)=\frac{2 i (\Theta_{\omega_1}^{\mathfrak{e}_6}(\mathbf{m}^{\mathfrak{e}_6},\tau)\!-\!\Theta_{\omega_6}^{\mathfrak{e}_6}(\mathbf{m}^{\mathfrak{e}_6},\tau))}{\eta(\tau)^{16}},
\end{equation}
where the level 1 $\mathfrak{e}_6$ theta functions
\begin{eqnarray*}
\Theta_{\omega_{1,6}}^{\mathfrak{e}_6}(\mathbf{m}^{\mathfrak{e}_6},\tau)\!\!&=&\!\!\!\!\sum_{w\in \Delta^{\mathfrak{e}_6}+\omega_{1,6}}\!\!\!\!\!\!\!\!\!\exp\left(\pi i \tau (w,w) +2\pi i (w,\mathbf{m}^{\mathfrak{e}_6}) \right)\!
\end{eqnarray*}
have the following $q$-series expansions:
\begin{eqnarray*}
\frac{\Theta^{\omega_1}_{\mathfrak{e}_6}(\mathbf{m},\tau)}{q^{11/24}\eta(\tau)^5}\!\!&=&\!\!\chi^{\mathfrak{e}_6}_{\mathbf{27}}\!+\!q\,\chi^{\mathfrak{e}_6}_{\mathbf{351}}\!+\!q^2(\chi^{\mathfrak{e}_6}_{\mathbf{1728}}\!+\!\chi^{\mathfrak{e}_6}_{\mathbf{351'}})\!+\!\mathcal{O}(q^3),\\
\frac{\Theta^{\omega_6}_{\mathfrak{e}_6}(\mathbf{m},\tau)}{q^{11/24}\eta(\tau)^5}\!\!&=&\!\!\chi^{\mathfrak{e}_6}_{\mathbf{\overline{27}}}\!+\!q\,\chi^{\mathfrak{e}_6}_{\mathbf{\overline{351}}}\!+\!q^2(\chi^{\mathfrak{e}_6}_{\mathbf{\overline{1728}}}\!+\!\chi^{\mathfrak{e}_6}_{\mathbf{\overline{351'}}})\!+\!\mathcal{O}(q^3).
\end{eqnarray*}
 In terms of $\mathfrak{d}_5\oplus {\mathfrak u}_1$ fugacities,
\begin{eqnarray}
\Theta_{\omega_1}^{\mathfrak{e}_6}(\mathbf{m},\tau)=&\!\!\!\!\frac{q^{1/6}}{2}\displaystyle{\sum_{k=1}^4}\sigma_k e^{-2\pi i \mu}\theta_k(3\mu\!-\!\tau,3\tau)\!\prod_{j=1}^5\theta_k(\mu_j,\tau),\nonumber\\
\Theta_{\omega_6}^{\mathfrak{e}_6}(\mathbf{m},\tau)=&\!\!\!\!\frac{q^{1/6}}{2}\displaystyle{\sum_{k=1}^4}\sigma_k e^{2\pi i \mu}\theta_k(3\mu\!+\!\tau,3\tau)\!\prod_{j=1}^5\theta_k(\mu_j,\tau),\nonumber\\
\label{eq:thetae616}
\end{eqnarray}
where $-\sigma_{1}=\sigma_{2}=\sigma_{3} =-\sigma_{4}=1$, $\mu= -2m^{\mathfrak{u}_1}$, and
\begin{eqnarray*}
\mu_1\!\!&=&\!\!m^{\mathfrak{d}_5}_1\!\!+\!\!m^{\mathfrak{d}_5}_2\!\!+\!\!m^{\mathfrak{d}_5}_3\!\!+\!\!\frac{1}{2}m^{\mathfrak{d}_5}_4\!\!+\!\!\frac{1}{2}m^{\mathfrak{d}_5}_5,\\
\mu_2\!\!&=&\!\!m^{\mathfrak{d}_5}_2\!\!+\!\!m^{\mathfrak{d}_5}_3\!\!+\!\!\frac{1}{2}m^{\mathfrak{d}_5}_4\!\!+\!\!\frac{1}{2}m^{\mathfrak{d}_5}_5,\\
\mu_3\!\!&=&\!\!m^{\mathfrak{d}_5}_3\!\!+\!\!\frac{1}{2}m^{\mathfrak{d}_5}_4\!\!+\!\!\frac{1}{2}m^{\mathfrak{d}_5}_5,\\
\mu_4\!\!&=&\!\!\frac{1}{2}m^{\mathfrak{d}_5}_4\!\!+\!\!\frac{1}{2}m^{\mathfrak{d}_5}_5\!,\, \mu_5 \!=\! -\frac{1}{2}m^{\mathfrak{d}_5}_4\!\!+\!\!\frac{1}{2}m^{\mathfrak{d}_5}_5
\end{eqnarray*}
are $\mathfrak{d}_5$ fugacities expressed in the orthogonal basis.

\bibliography{PSE6}

\end{document}